\documentclass[conference]{IEEEtran}

\usepackage{algorithm}
\usepackage{algpseudocode}
\usepackage{amsmath}
\usepackage{amssymb}
\usepackage{graphicx}
\usepackage{lettrine}
\usepackage{multirow}
\usepackage{subfigure}
\usepackage{threeparttable}

\begin{document}

\title{Base Station Switching Problem for Green Cellular Networks with Social Spider Algorithm}

\author{James J.Q. Yu,
       \textit{Student Member, IEEE} and
       Victor O.K. Li,
       \textit{Fellow, IEEE}\\
       Department of Electrical and Electronic Engineering\\
       The University of Hong Kong\\
       Email: \{jqyu, vli\}@eee.hku.hk}
       
\maketitle
\pagestyle{empty}

\begin{abstract}
With the recent explosion in mobile data, the energy consumption and carbon footprint of the mobile communications industry is rapidly increasing. It is critical to develop more energy-efficient systems in order to reduce the potential harmful effects to the environment. One potential strategy is to switch off some of the under-utilized base stations during off-peak hours. In this paper, we propose a binary Social Spider Algorithm to give guidelines for selecting base stations to switch off. In our implementation, we use a penalty function to formulate the problem and manage to bypass the large number of constraints in the original optimization problem. We adopt several randomly generated cellular networks for simulation and the results indicate that our algorithm can generate superior performance.
\end{abstract}

\begin{IEEEkeywords}
Green cellular network, base station switching, social spider algorithm, evolutionary computation, swarm intelligence.
\end{IEEEkeywords}

\section{Introduction}
With the growing concern on the global climate change, it becomes critical to develop energy-efficient systems in all industries. In particular, the information and communication technology (ICT) sector was estimated to contribute to more than 830 million tons of carbon dioxide emission in 2013, which is approximately 2\% of the global carbon dioxide emission, and this number is expected to double by the year 2020 \cite{2013CEETAnnualReport}. With the fast growing mobile data communication demand, the wireless cellular networks are playing a more important role in ICT sector than ever before. Thus a promising approach to reduce the global green house gas emission is to reduce the energy consumption of the cellular networks. It is also very important from the economical perspective of network operators as a significant portion of the operational expenditure is due to electricity consumption.

Designing green cellular networks, especially green base stations, is a recent hot research topic. There are at least two mainstream approaches. With the development of smart grid technology, one approach is to utilize renewable energy for base station operation \cite{TaoAnsari2013OptimizingGreenEnergy}. Another approach is to manage the operation profile of the base stations. As the network operators need to deploy their base stations to support the peak mobile data traffic, it is inevitable that during a major portion of the day a large number of the base stations are under-utilized. This phenomenon can be observed from real mobile data usage profile \cite{OhSonKrishnamachari2013DynamicBaseStation}. Unfortunately, the energy consumption of an idle base station is nearly the same as one under full load \cite{CorreiaZellerBlumeFerlingJadingGodorAugerPerre2010Challengesandenabling}. So an under-utilized base station must be switched off in order to save energy, and there naturally rises a problem: Which base stations should be switched off to save the most amount of energy while maintaining adequate service for end users?

Such energy-efficient design of wireless cellular networks has attracted significant attention recently. However, most of the existing efforts focus on the formulation of the problem \cite{OhSonKrishnamachari2013DynamicBaseStation}\cite{MarsanChiaraviglioCiulloMeo2009OptimalEnergySavings}. This paper focuses on designing an swarm intelligence algorithm based on the Social Spider Algorithm (SSA) \cite{YuLi2013SocialSpiderAlgorithm}. In this paper, we concentrate on the modifications on the original problem to make it solvable with metaheuristics, and the implementation of a metaheuristic.

The rest of the paper is organized as follows. In Section \ref{sec:background} we briefly introduce the current progress on green cellular network design. In Section \ref{sec:formulation} we elaborate on the original optimization problem and our modifications. Section \ref{sec:algorithm} presents the detailed implementation of our proposed algorithm. Section \ref{sec:simr} introduces the random network generation protocol, and the simulation results are presented with analysis and discussion. Finally we conclude in Section \ref{sec:conclusion} with potential future research.

\section{Background}\label{sec:background}
Green cellular network design is a recent hot research topic, and much work has been carried out. Chiaraviglio \textit{et al.} initially proposed the idea of dynamic base station operation based on the data traffic profile in \cite{ChiaraviglioCiulloMeoMarsan2009Energyefficientmanagement}. Later they extended their work and set up an analytical model to control the base station switching profile \cite{MarsanChiaraviglioCiulloMeo2009OptimalEnergySavings}. Oh and Krishnamachari also proposed a similar model in \cite{OhKrishnamachari2010EnergySavingsthrough}, and devised a simple base station switching policy. They further developed a complete system model for this problem and proposed a distributed algorithm to solve it \cite{OhSonKrishnamachari2013DynamicBaseStation}. In this work, both the algorithm design and the practical implementation, including a distributed base station switching protocols were elaborated, and the preliminary simulation results indicate a satisfactory energy-saving performance. All the above work focus on base station operations in the same type of networks, i.e. macro cellular networks. There are also some researchers concentrating on the cooperative operation considering network sharing. Fehske \textit{et al.} investigated the possibility of deploying small and low power base stations alongside conventional ones from a deployment perspective in \cite{FehskeRichterFettweis2009EnergyEfficiencyImprovements}. Marsan and Meo evaluated the energy savings achieved with the energy-aware cooperative management of the cellular access networks of different operators over the same area in \cite{MarsanMeo2010Energyefficientmanagement}. In \cite{RostFettweis2010TransmissionComputationEnergy}, Rost and Fettweis discussed the energy-efficient operation based on the cooperative transmission in multi-hop systems. Niu \textit{et al.} introduced the concept of cell zooming for energy savings. In their implementation, the cell size is adaptively adjusted according to different control variables.

\section{Problem Formulation}\label{sec:formulation}
In this paper, we adopt the system model formulated in \cite{OhSonKrishnamachari2013DynamicBaseStation} as it is a simple yet generalized wireless cellular network model. In this model, the system load is formulated as follows:
\begin{equation}
\rho_b(t)=\int_{A_b}\frac{\gamma(x,t)}{r_b(x,t)}dx,
\end{equation}
where $\rho_b(t)$ represents the system load of base station $b$ at time $t$, $A_b$ represents $b$'s coverage, $\gamma(x,t)$ represents the traffic load of user equipment $x$ at $t$, and $r_b(x,t)$ represents the service rate of $x$ from $b$ at $t$. This system load stands for the fraction of time needed to serve all the data transmission load in the base station's coverage. For simplicity, we omit the time variable in the following notations.

With this system model, we further formulate our Base Station Switching Problem (BSSP) as follows. We consider that the system load for a wireless cellular network remains constant in a short time interval. An optimization problem may be  formulated as follows:
\begin{equation}\label{eqn:fitness}
\begin{aligned}
\min&\hspace{1em}\sum_{b\in B}a_b\\
s.t.&\hspace{1em}0\leq\rho_b\leq\overline{\rho_b},\hspace{1em}\forall b\in B
\end{aligned}
,
\end{equation}
where $B$ is the collection of all available base stations, $a_b\in\{0,1\}$ is the active indicator of $b$, and $\overline{\rho_b}$ is the maximum allowed system load for $b$. When a base station $b$ is switched off, its traffic load will be handled by all its active neighboring base stations $N_b = \{N_{b,1},N_{b,2},\cdots,N_{b,n}\}$, and we use $\rho_{b\rightarrow i}$ to denote the amount of transferred traffic load from $b$ to the $i$-th neighbor base station in $N_b$.

From (\ref{eqn:fitness}) we can see the fitness evaluation function is very simple, but the rigid constraints may potentially obstruct the generation of feasible solutions to the problem. In order to alleviate the effort in designing an algorithm that can easily generate solutions satisfying all the constraints, we transform the constraint in (\ref{eqn:fitness}) into a penalty function. The new fitness function is described as follows:
\begin{equation}\label{eqn:newfitness}
\min\hspace{1em}\sum_{b\in B}(a_b + |B|\times P_b),
\end{equation}
where $P_b$ is the penalty value for $b$. In order to define $P_b$, we first analyze the possible scenarios a base station may come across during the switching operation:
\begin{enumerate}
\item $b$ is switched off, and $|N_b|=0$. In this case, both $b$ and all its neighboring base stations are switched off and the traffic load originally handled by $b$ cannot be served anymore. $P_b$ is defined as $1+\rho_b$ for all such base stations.
\item $b$ is switched off, and $|N_b|>0$. In this case, $b$ is switched off, but it has some active neighboring base stations. The traffic load of $b$ is handed over to $N_b$ and the transferred load to $N_{b,i}$ is $\rho_{b\rightarrow i}$. $P_b$ is defined as $0$ for all such base stations.
\item $b$ remains active, and a number of its neighboring base stations are switched off. We use $S_b$ to denote the collection of the neighboring inactive base stations. Then the system load for $b$ is $\rho_b+\sum_{s\in S_b}\rho_{s\rightarrow b}$. If this new system load is larger than $\overline{\rho_b}$, $P_b$ is defined as $1+\rho_b+\sum_{s\in S}\rho_{s\rightarrow b}-\overline{\rho_b}$. Otherwise, $P_b=0$ for all such base stations.
\end{enumerate}

This penalty function is designed according to the design principles that are critical for a successful penalty function stated in \cite{KhuriBackHeitkotter1994evolutionaryapproachto}:
\begin{enumerate}
\item The fitness values shall improve as the solutions approach feasible regions of the search space.
\item The unfeasible solutions are guaranteed to be assigned with fitness values inferior to the fitness value of the worst feasible solution.
\end{enumerate}
Thus, any solutions with fitness values larger than $|B|$ are unfeasible ones to the original optimization problem (\ref{eqn:fitness}).

It is noted that BSSP is NP-complete as it can be reduced from a vertex-covering problem, which has been shown to be NP-complete \cite{Karp1972Reducibilityamongcombinatorial}.

\section{Algorithm Design}\label{sec:algorithm}

In this section, we will first briefly introduce SSA. Then the detailed implementation of our proposed methodology will be presented.

\subsection{Social Spider Algorithm}

SSA is a recently proposed general-purpose swarm intelligence algorithm. It mimics the foraging behavior of the social spiders to perform optimization task. In SSA, the search space of the optimization problem is formulated as a hyper-dimensional spider web, and each position on the web represents a feasible solution. Besides the solution space, the spider web also serves as the transmission media of the vibrations generated by the spiders.

The spiders are the basic operating agents of SSA. Each spider holds a memory consisting of its current position on the web, the fitness value of its current position, and the vibration (position and intensity) it was following in the previous iteration. The first two pieces of information describe the characteristic of this spider, and the last one helps the algorithm guide the movement of this spider.

Based on biological observations, spiders are found to be extremely sensitive to vibrations. They can accurately sense the strength and the direction of vibrations, and can even separate different vibrations propagated on the same web\cite{Uetz1992Foragingstrategiesspiders}. SSA utilizes this characteristic of the spiders and established a vibration-based information-loss communication system among spiders. In SSA, a spider generates a new vibration whenever it moves to a different position from the previous one. The vibration will then propagate over the spider web and be sensed by others. By this means, the spiders share their personal information to form a collective social knowledge.

The vibrations are defined by two properties in SSA, namely, the source position and the source intensity. When a spider moves to a new position, it generates a vibration at its current position. We define the intensity of a vibration in the range $[0,+\infty)$, and the value of the intensity is calculated as follows:
\begin{equation}\label{eqn:intensity}
    I=
    \begin{cases}
    1/(C_{max} - f(s)) & \mathrm{for\:maximization} \\
    1/(f(s) - C_{min}) & \mathrm{for\:minimization} \\
    \end{cases}
    ,
\end{equation}
where $I$ is the intensity of the vibration at its source position, $f(s)$ is the fitness value of the source position, $C_{max}$ is a confidently large constant selected such that all possible fitness values of the maximization problem are smaller than $C_{max}$, and $C_{min}$ is a confidently small constant such that all possible fitness values of the minimization problem are larger than $C_{min}$.

(\ref{eqn:intensity}) provides the method of calculating the source intensity of the vibration. However, as a kind of energy, the vibration attenuates over time and distance during the process of propagation. This physical phenomenon is also taken into consideration in SSA from two aspects, i.e., time and distance. As a swarm intelligence algorithm, SSA performs the searching task in an iterative manner. The vibrations are attenuated with each iteration as follows:
\begin{equation}\label{eqn:vibratteniter}
    I(t+1)=I(t)\times r_a,
\end{equation} 
where $I(t)$ is the source vibration intensity at iteration $t$, and $r_a$ is a user-defined vibration attenuation parameter. This design can prevent the algorithm from pre-mature convergence as a non-decaying vibration on the web can potentially attracts all spiders to move continuously towards it, thus facilitating the exploitation searching behavior but obstructing the exploration of the whole search space.

Another attenuation factor besides the time is the propagation distance. In SSA, the vibration attenuation over distance is defined as follows:
\begin{equation}\label{eqn:vibrattendist}
    I(p)=I(s)\times\exp(-\frac{D(s,p)}{D_{max}\times r_a}),
\end{equation}
where $s$ and $p$ are the source position and receiving position of the vibration, respectively, $I(s)$ is the vibration intensity at $s$, $D(s,p)$ is the distance between $s$ and $p$, and $D_{max}$ is the maximum possible distance between any two positions on the web. We usually employ the Manhattan distance to reduce the computational time.

There are three phases in SSA, namely, initialization, iteration, and final. In each run of SSA, we start with the initialization phase where the objective function, search space, and the optimization parameters of SSA are initialized. Then a random population of spiders are generated and placed on the spider web, i.e., search space. The positions of these spiders are randomly generated in the search space, and the initial vibration each spider is following has a position of the spider's current position, and an intensity of zero. This finishes the initialization phase and SSA proceeds to the iteration phase.

In the iteration phase, the algorithm performs the optimization task in an iterative manner. A number of iterations are executed. In each iteration, the algorithm first evaluates the fitness values of all spiders on the web and attenuates the vibrations in the previous iteration. Then the spiders generate vibrations using (\ref{eqn:intensity}) at their current positions. The vibrations then propagates using (\ref{eqn:vibrattendist}) over the spider web. After the propagation, each spider will receive a number of different vibrations from all directions. Upon the receipt of these attenuated vibrations, each spider chooses one vibration with the largest intensity, i.e., the strongest vibration $v_{best}$. $v_{best}$ is then compared with the stored vibration this spider followed in the previous iteration $v_{prev}$. The one with a larger intensity is kept and saved as $v_{best}$. The algorithm then manipulate the position of the spiders as follows:
\begin{equation}\label{eqn:walk}
    \boldsymbol{P}_s(t+1)=\boldsymbol{P}_s(t)+(\boldsymbol{P}_{best}(t)-\boldsymbol{P}_s(t))\odot(\boldsymbol{1}-\boldsymbol{R}\odot\boldsymbol{R}),
\end{equation}
where $\boldsymbol{P}_s(t)$ is the position of spider $s$ at iteration $t$, and $\odot$ denotes element-wise multiplication. $\boldsymbol{P}_{best}$ is the vibration source position of the best vibration $v_{best}$. $\boldsymbol{R}$ is a vector of random numbers generated from zero to one uniformly, whose length is the number of dimensions of the problem, and $\boldsymbol{1}$ is a vector of ones of the same length as $\boldsymbol{R}$.

After the manipulation of spider positions, an artificial spider jump away process is introduced in order to prevent the algorithm from getting stuck in the local optimums. Each spider has a small possibility of being re-assigned with a new random position in the search space. And this ends one iteration.

The iteration phase loops until the stopping criteria is matched. The algorithm then outputs the best solution with the fitness value. The above three phases constitute SSA and interested readers can refer to \cite{YuLi2013SocialSpiderAlgorithm} for more details.

\subsection{The Proposed Implementation}
SSA was initially proposed to solve global optimization problems, i.e., the solution space is continuous. So we make several modifications to adapt it to solve BSSP, which is a discrete problem.

\subsubsection{Encoding Scheme}
In this work, we adopt the classical encoding scheme for this kind of problem. We use a vector of 0's and 1's to represent the off/on state of each available base station.

\subsubsection{Spider Following Scheme}
In SSA, we use (\ref{eqn:walk}) to manipulate the movement of spiders. However this equation is designed for continuous optimization. So we devise a new spider following scheme to replace (\ref{eqn:walk}) as follows:
\begin{equation}\label{eqn:newwalk}
    \boldsymbol{P}_s(t+1)=\boldsymbol{P}_s(t)+round((\boldsymbol{P}_{best}(t)-\boldsymbol{P}_s(t))\odot\boldsymbol{R}),
\end{equation}
where $round()$ is the rounding function. (\ref{eqn:newwalk}) will first determine the dimensions at which $\boldsymbol{P}_{best}(t)$ and $\boldsymbol{P}_s(t)$ are different. Then all different dimensions have a probability of $0.5$ to change.

\subsubsection{Spider Jump Away Scheme}
Besides the spider following scheme, another scheme that can change the position of the spiders is the spider jump away scheme. In SSA, the jump away scheme operates at the spider level, where each selected spider is assigned with a new random position. But in our implementation for solving BSSP, the jump away scheme operates on the dimension level. Right after the spider following scheme has manipulated the position of each spider, each dimension of a spider position has a probability of $1/|B|$ to change from zero to one or one to zero, i.e.,
\begin{equation}\label{eqn:newjump}
    \boldsymbol{P}_s(t+1)=|step(\boldsymbol{R})-\boldsymbol{P}_s(t)|,
\end{equation}
where $step(r)=1$ if $r<1/|B|$, otherwise $step(r)=0$.

These are the modifications we made to adapt SSA to solve BSSP. Algorithm \ref{alg:ssa} is the pseudo-code of SSA for BSSP

\begin{algorithm}
\caption{\sc{Social Spider Algorithm for Base Station Switching Problem}}
  \begin{algorithmic}[1]
  \State Assign values to the parameters of SSA.
  \State Create the population of spiders $pop$ and assign memory for them.
  \State Initialize $v_{prev}$ for each spider.
  \While {stopping criteria not met}
    \For {\textbf{each} spider $s$ in $pop$}
      \State Evaluate the fitness value of $s$.
      \State Attenuate the intensity of $v_{prev}$.
      \State Generate a vibration at the position of $s$.
    \EndFor
    \For {\textbf{each} spider $s$ in $pop$}
      \State Calculate the intensity of the vibrations $V$ \par\hspace{0.7em} generated by other spiders.
      \State Select the strongest vibration $v_{best}$ from $V$.
      \If {The intensity of $v_{best}$ is smaller than $v_{prev}$}
        \State Store $v_{prev}$ as $v_{best}$.
      \EndIf
      \State Perform the spider following move towards $v_{best}$.
      \State Perform the spider jump away modification on \par\hspace{0.7em} each spider.
      \State Store $v_{best}$ as $v_{prev}$.
    \EndFor
  \EndWhile
  \State Output the best solution found.
  \end{algorithmic}
  \label{alg:ssa}
\end{algorithm}

\section{Experimental Setting and Simulation Results}\label{sec:simr}

In order to evaluate the performance of saving energy using our proposed algorithm, we performed a series of simulations over some randomly generated instances.

\subsection{Testing Instance}

In our simulation, we consider the 3G network topologies consisting of 20, 40, and 60 base stations in the area of $10 \times 10$ km$^2$. The instances are generated as follows. We first randomly deploy the base stations to different positions in the area. Then we start from the first base station and determine the number of its neighboring base stations using
\begin{equation}\label{eqn:neighbornumber}
    |Neigh_b|=|N_b+S_b|=Pois(\lambda-2)+2,
\end{equation}
where $Pois(\lambda)$ is the Poisson distribution and $\lambda$ is the expected number of neighboring base stations. Each base station may have a different number of neighborhoods, and this design guarantees that all base stations will have at least two neighborhoods.

After determining the number of neighborhoods for each base stations, we establish the neighboring links starting from the first base station to the last one. In this step we first check the existing links connected to the current base station. If the number is already larger than $|Neigh_b|$, we go on to the next base station. Otherwise we establish a neighboring link between the current base station and its nearest base station that is not linked yet. This substep will iterate until enough links are established. The pseudo-code for this random instance generation algorithm is presented in Algorithm \ref{alg:random}.

\begin{algorithm}
\caption{\sc{Random Instance Generation Algorithm}}
  \begin{algorithmic}[1]
  \For {\textbf{each} base station $b$ in $B$}
    \State Assign a random position in the area for $b$.
  \EndFor
  \For {\textbf{each} base station $b$ in $B$}
    \State Determine the number of $b$'s neighboring base stations \par $|Neigh_b|$.
  \EndFor
  \For {\textbf{each} base station $b$ in $B$}
    \State Find the number of $b$'s established neighboring links \par $n$.
    \If {$n\geq|Neigh_b|$}
      \State \textbf{continue}
    \Else
      \State Find the $|Neigh_b|-n$ nearest base stations to $b$ \par\hspace{0.7em} that not yet linked with $b$.
      \State Establish neighboring links between these base \par\hspace{1.1em}stations with $b$.
    \EndIf
  \EndFor
  \State Output the best solution found.
  \end{algorithmic}
  \label{alg:random}
\end{algorithm}

\begin{figure}
  \center
  \subfigure[20.3.A]{
   \includegraphics[width=0.27\linewidth]{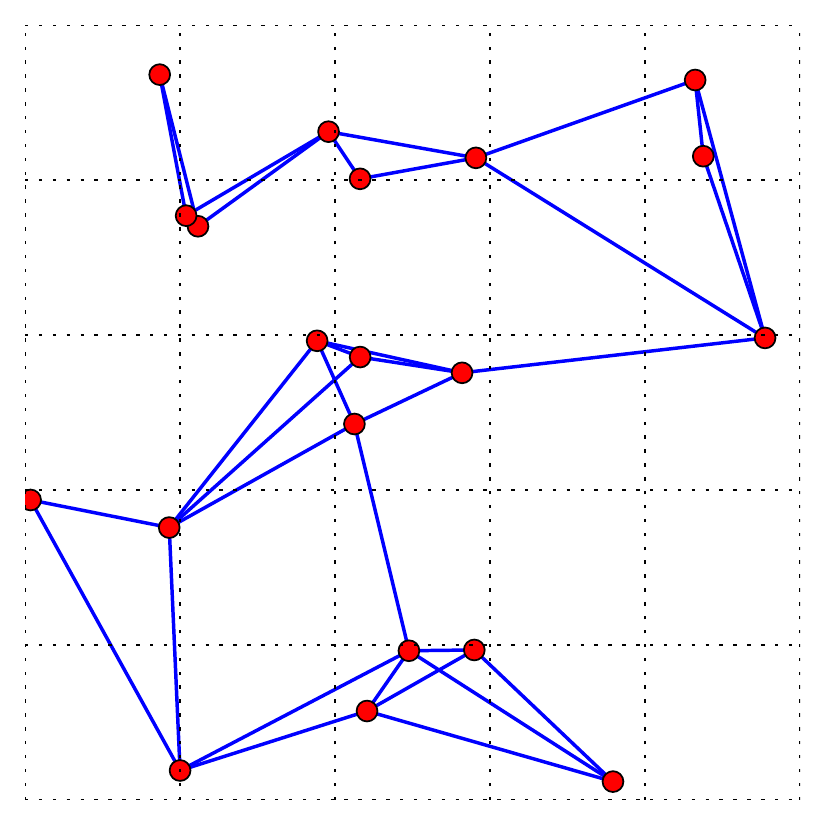}
 }
 \subfigure[20.3.B]{
   \includegraphics[width=0.27\linewidth]{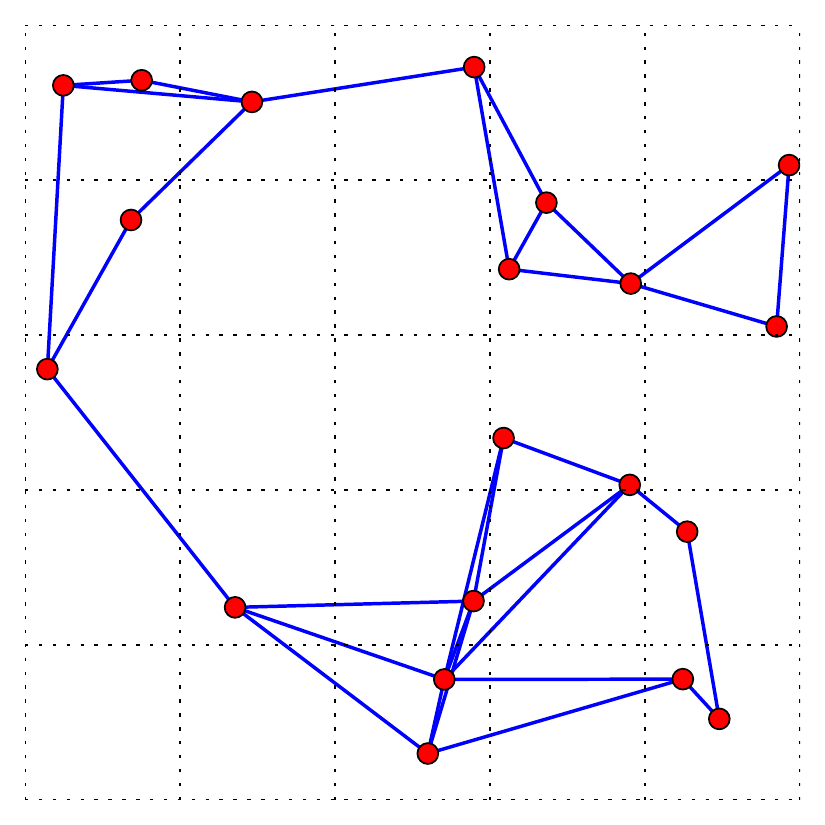}
 }
 \subfigure[20.3.C]{
   \includegraphics[width=0.27\linewidth]{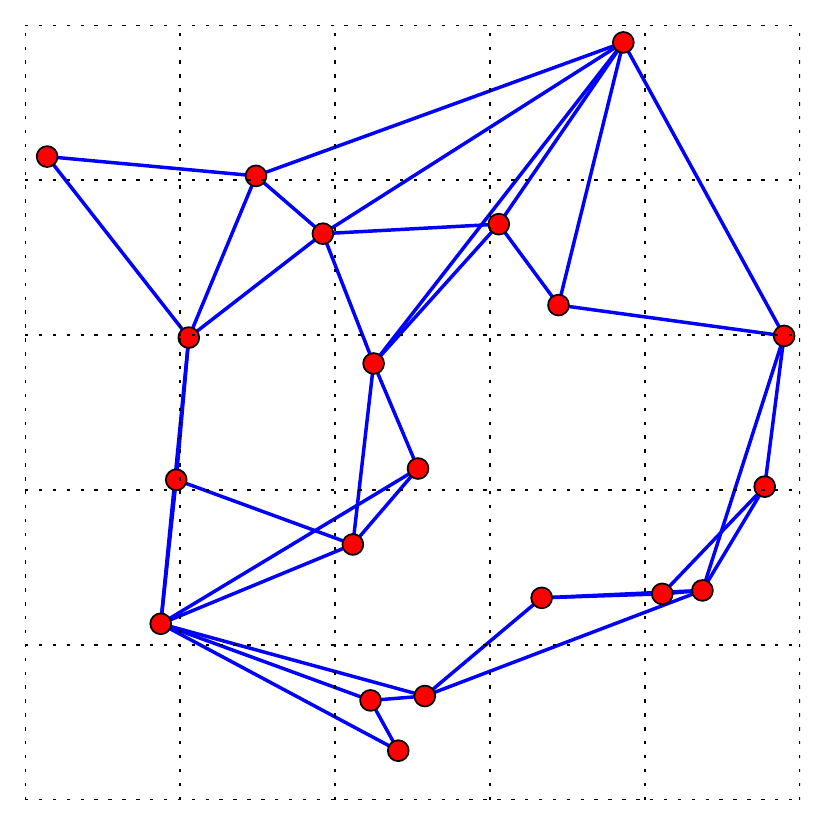}
 }
 
  \subfigure[40.4.A]{
   \includegraphics[width=0.27\linewidth]{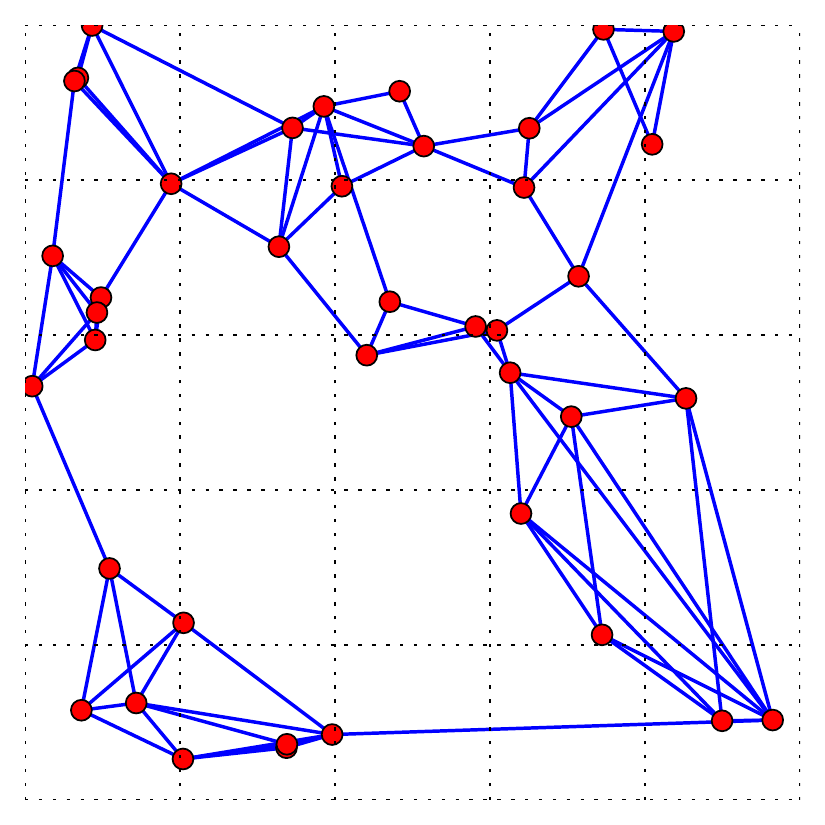}
 }
 \subfigure[40.4.B]{
   \includegraphics[width=0.27\linewidth]{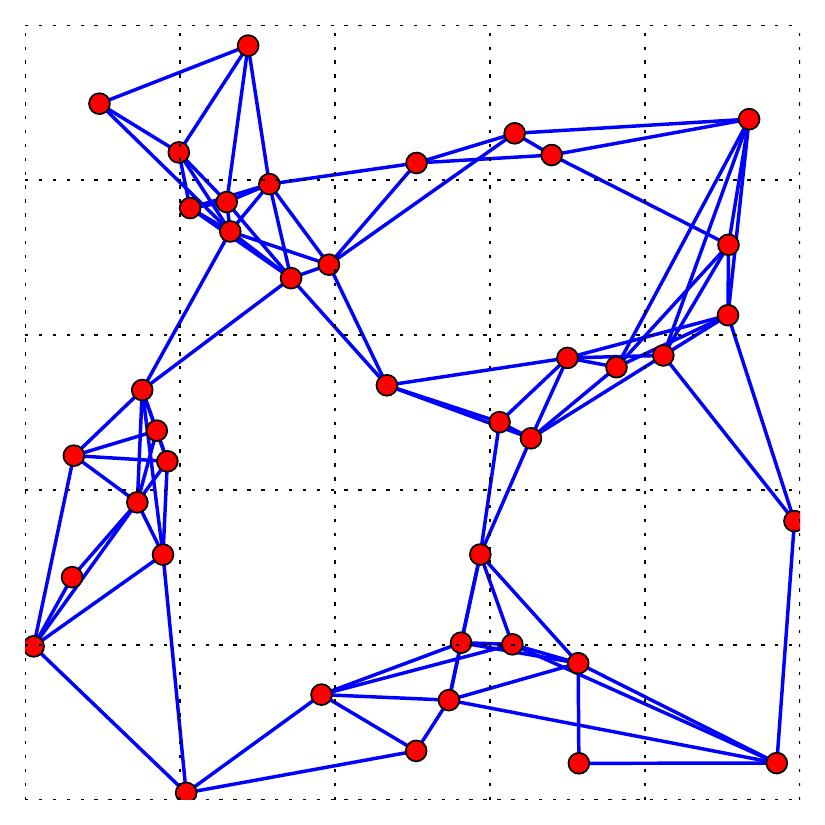}
 }
 \subfigure[40.4.C]{
   \includegraphics[width=0.27\linewidth]{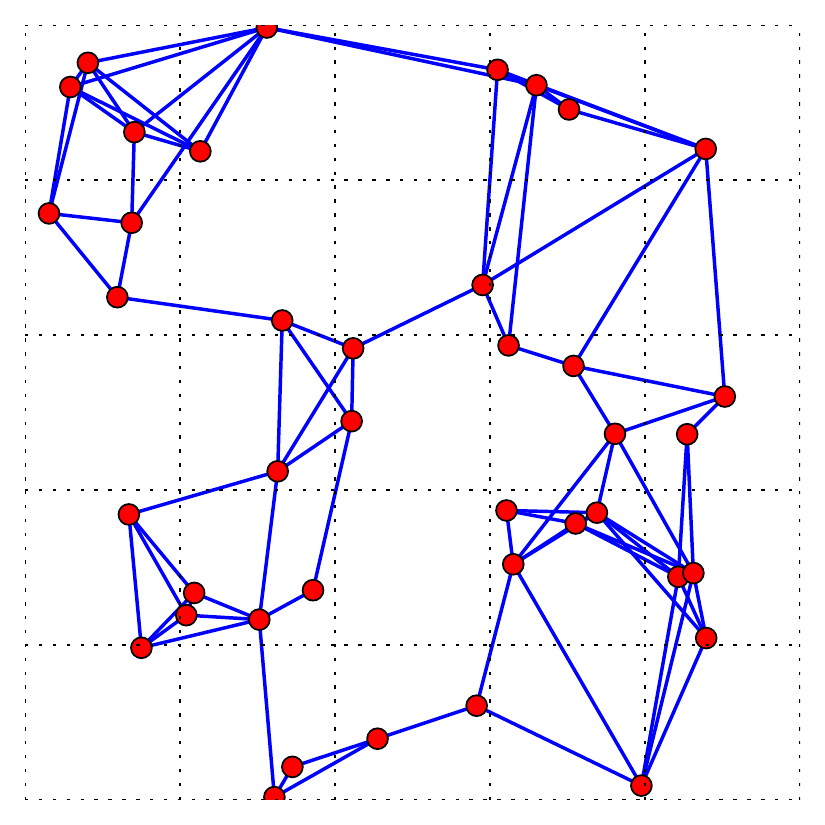}
 }
 
  \subfigure[60.5.A]{
   \includegraphics[width=0.27\linewidth]{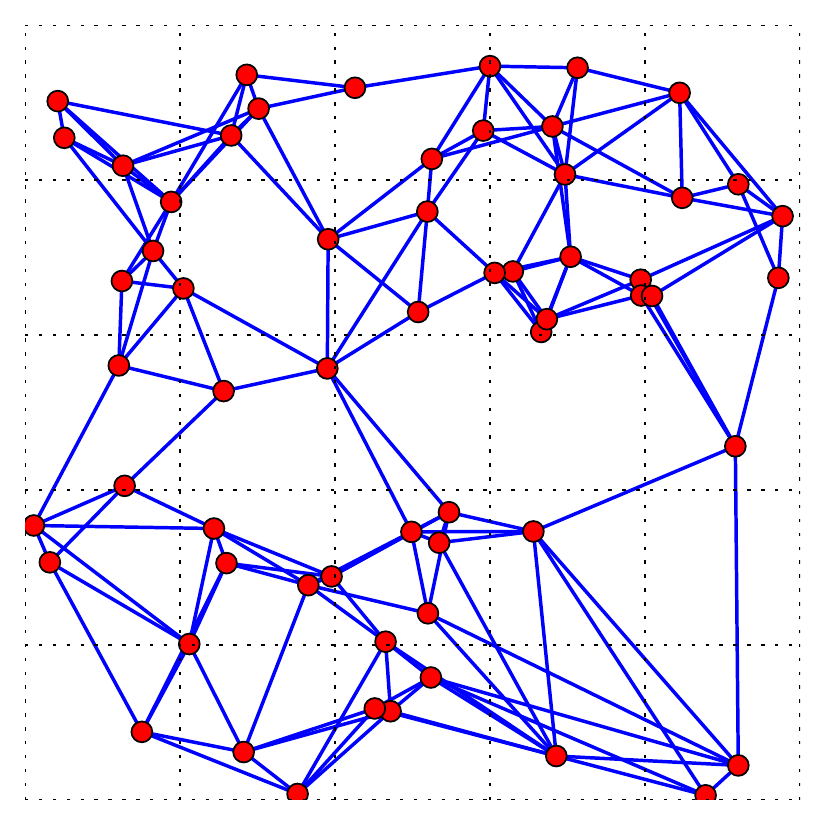}
 }
 \subfigure[60.5.B]{
   \includegraphics[width=0.27\linewidth]{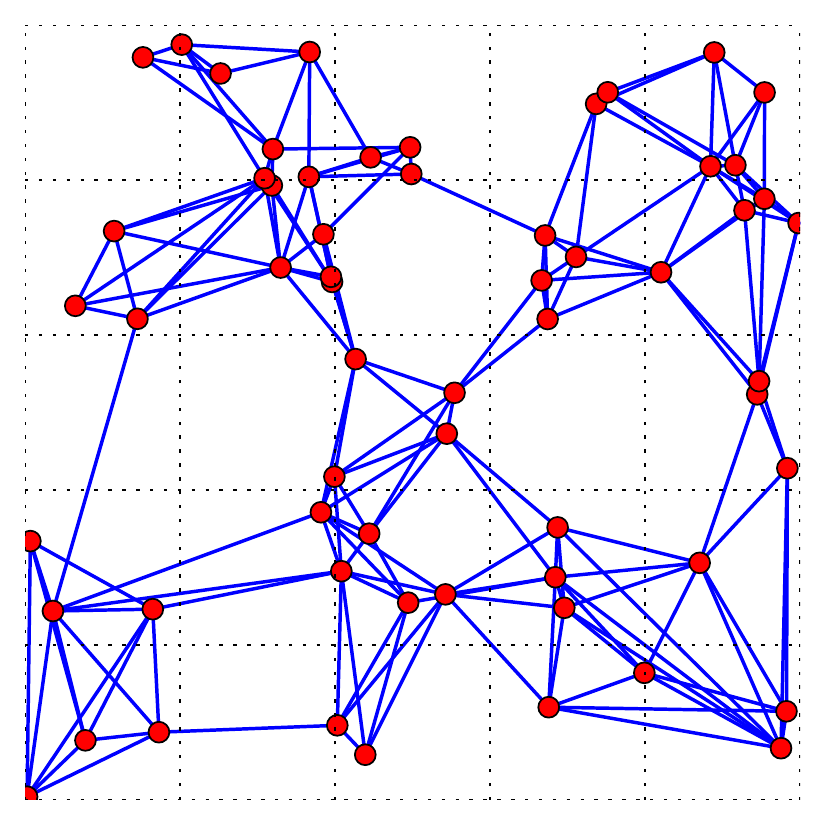}
 }
 \subfigure[60.5.C]{
   \includegraphics[width=0.27\linewidth]{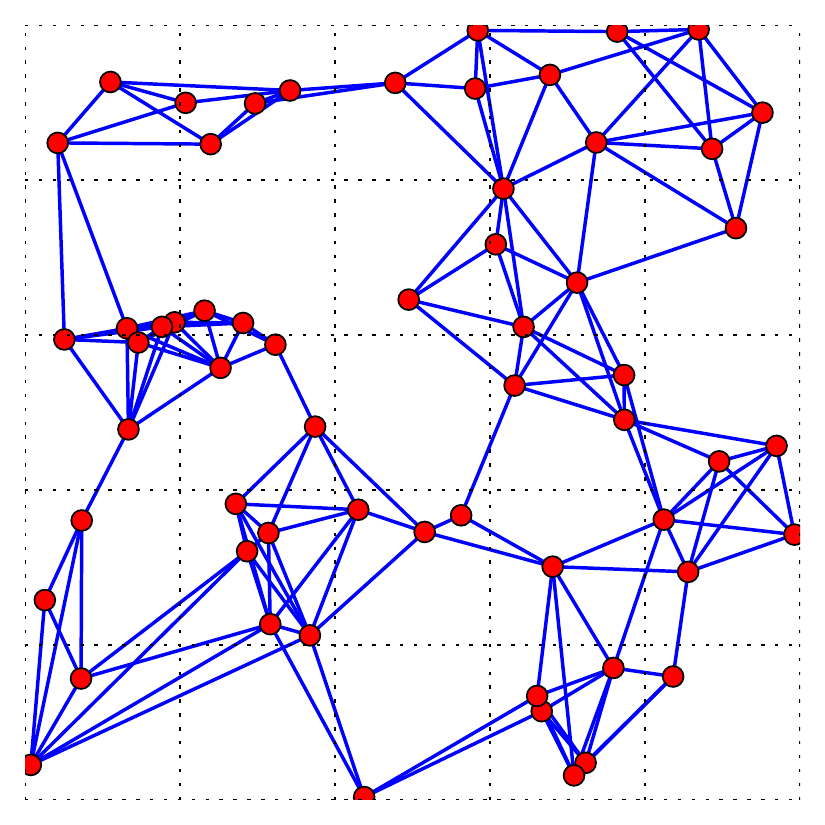}
 }
  \caption{Plot of Randomly Generated Topology Instances with Different Parameters}
  \label{fig:instance}
\end{figure}

Using this method we generated 9 instances using the instance parameter $(|B|,\lambda)\in[(20,3),(40,4),(60,6)]$, and 3 different instances are generated for each instance parameter pair. These instances are shown in Fig. \ref{fig:instance}, where the dots are the base stations and the base stations connected with lines are neighboring base stations. The first two digits in the names of the instances are $|B|$ and $\lambda$, respectively, and the last letter stands for different instances with the same parameters.

\subsection{Experimental Setting}
To determine which base stations to switch off, we need some feedback information from all the base stations. However these feedback information may reduce the overall performance and sometimes is invalid. So in our simulation we use an approximation method to reduce the reliability of these external and internal information. We make two assumptions on the random instances generated:
\begin{enumerate}
\item The system load for each base station is equal to the overall system load: $\rho_b\equiv\rho_s, \forall b\in B$. Here $\rho_s$ is the overall system load.
\item The transferred traffic load from one base station to all its neighbor base stations are equal: $\rho_{b\rightarrow i}\equiv\rho_b/|B|$
\end{enumerate}
With these two assumptions we minimize the impact of all additional information and focus on the optimization performance of our proposed algorithm.

We compare the performance of our proposed SSA for BSSP with the SWitching-on/off based Energy Saving algorithm (SWES) \cite{OhSonKrishnamachari2013DynamicBaseStation} whose system model was adopted in our problem formulation. In particular, the SWES$_{(0,0)}$ algorithm is employed for performance comparison as no additional information (external or internal) is required for this algorithm, thus satisfying our assumptions. For SSA, we use the number of base stations as the size of the population, and the vibration attenuation rate is set to 0.9. As the spider jump away scheme is modified, we do not need the jump away rate as elaborated in \cite{YuLi2013SocialSpiderAlgorithm}. The maximum iteration number is set to 500, so for 20-base-station instances the fitness function is evaluated for 10 000 times. $\overline{\rho_b}$ is set to 0.6 for all base stations in all the instances and we test the performance of the compared algorithms with $\rho_s\in[0.05,0.1,0.15,0.2,0.25,0.3,0.35,0.4,0.45,0.5,0.55]$ for each instance.

\subsection{Simulation Results}

\begin{figure}
  \center
  \subfigure[Mean results of 20 base stations with $\lambda=3$]{
   \includegraphics[width=0.95\linewidth]{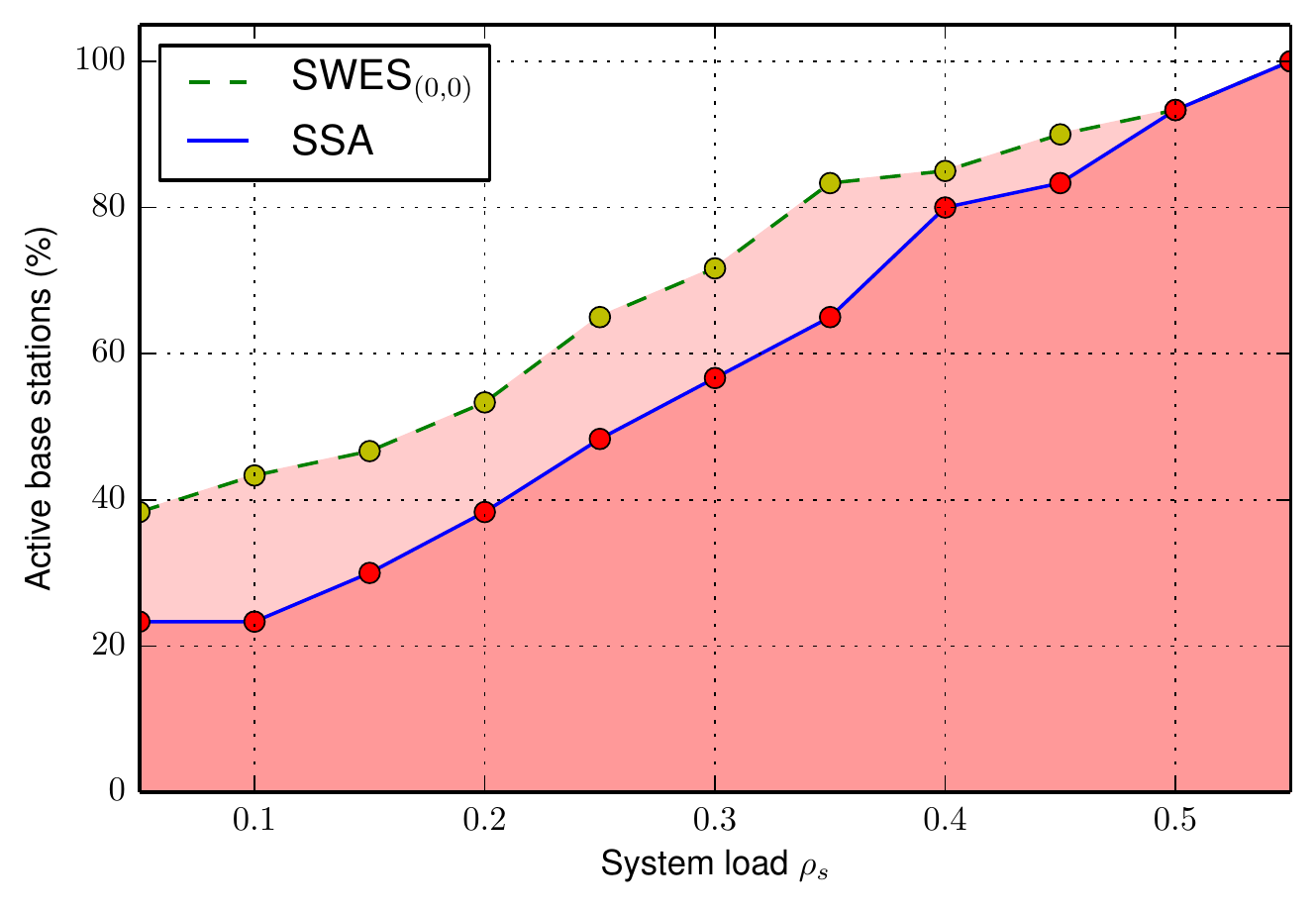}
 }
  \subfigure[Mean results of 40 base stations with $\lambda=4$]{
   \includegraphics[width=0.95\linewidth]{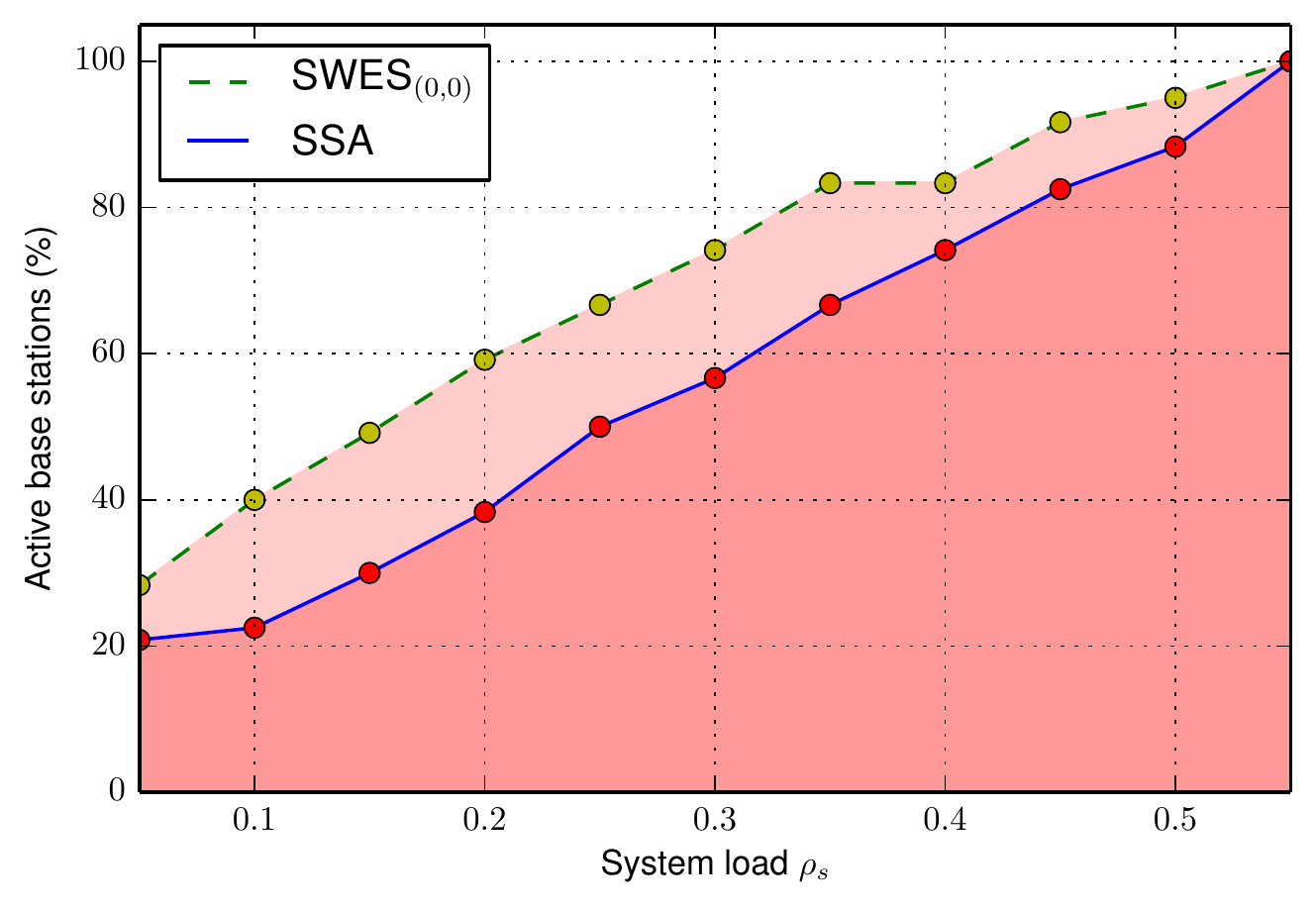}
 }
  \subfigure[Mean results of 60 base stations with $\lambda=5$]{
   \includegraphics[width=0.95\linewidth]{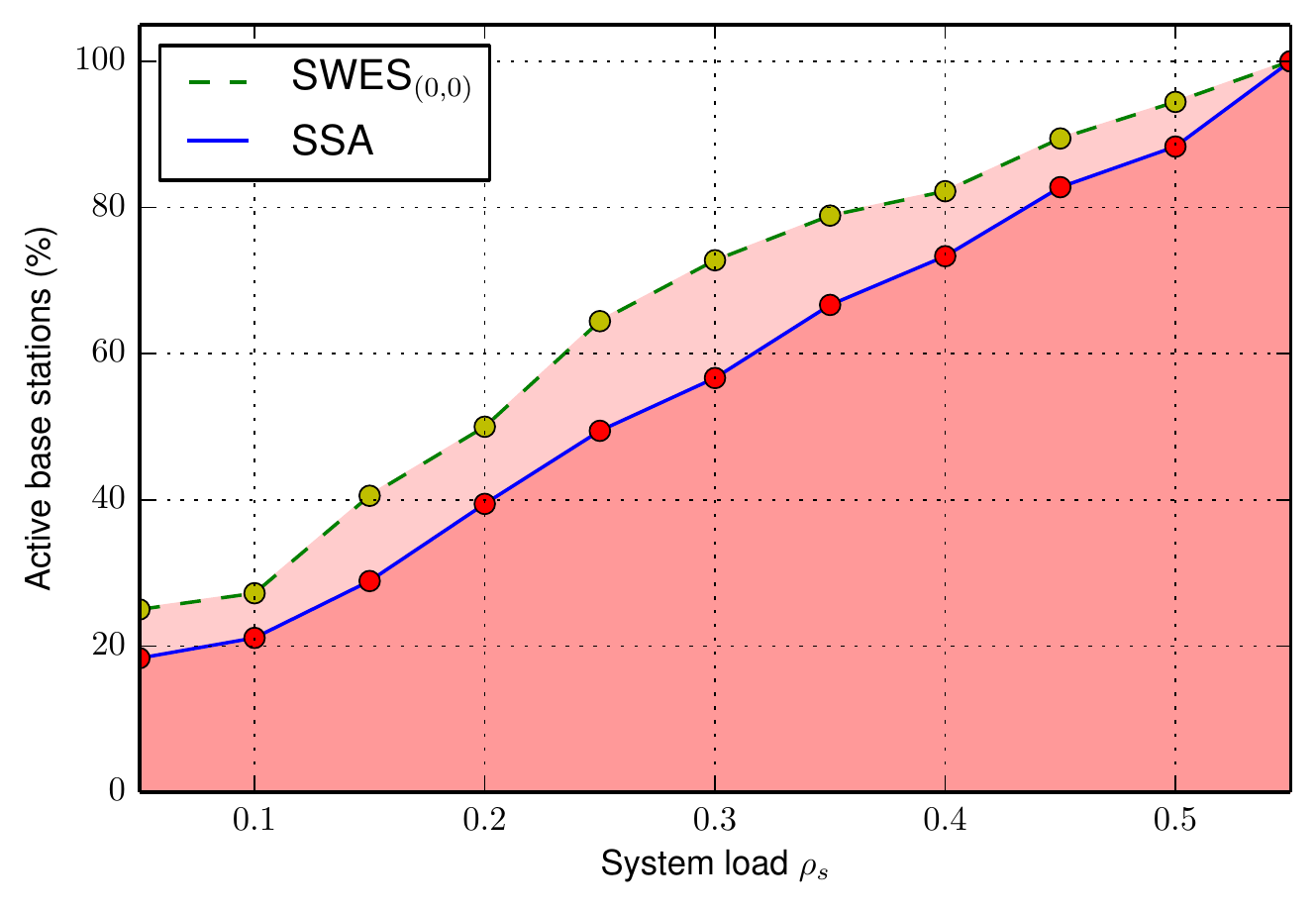}
 }
  \caption{Comparison of Active Base Stations}
  \label{fig:result}
\end{figure}

Fig. \ref{fig:result} illustrates the simulation results of our generated random topology instances with SSA for BSSP and SWES$_{(0,0)}$. Each dot in the figure is the mean value of the number of active base stations in all three instances with the corresponding same set of parameters. From the simulation results we have the following observations:

\begin{enumerate}
\item SSA for BSSP always outperforms SWES$_{(0,0)}$ in all three set of simulations. Out of 33 data points, SSA performs better than SWES$_{(0,0)}$ in 29, and the remaining 4 comparisons end up with draws.

\item Both SSA for BSSP and SWES$_{(0,0)}$ are efficient in terms of saving energy. Assume the traffic profile is increasing and decreasing evenly during the day, SSA can save 42.5\% of the total energy and SWES$_{(0,0)}$ can save 31.3\%. These numbers are generated by averaging all data points indicated in Fig. \ref{fig:result}.

\item The advantage of SSA for BSSP over SWES$_{(0,0)}$ is more significant when the system load is relatively low. A potential reason is that when the system load is close to the maximum allowed system load, the number of base stations that can potentially be switched off is very small. This means that the number of feasible solutions to (\ref{eqn:fitness}) is very small and a heuristic like SWES$_{(0,0)}$ can also find a good solution to the problem. In the extreme case of $\rho=0.55$, no base station can be switched off and the performance of the two compared algorithm naturally becomes the same.
\end{enumerate}

\subsection{Impact of $\lambda$ on Energy Saving Performance}
\begin{figure}
  \center
  \subfigure[40.3]{
   \includegraphics[width=0.27\linewidth]{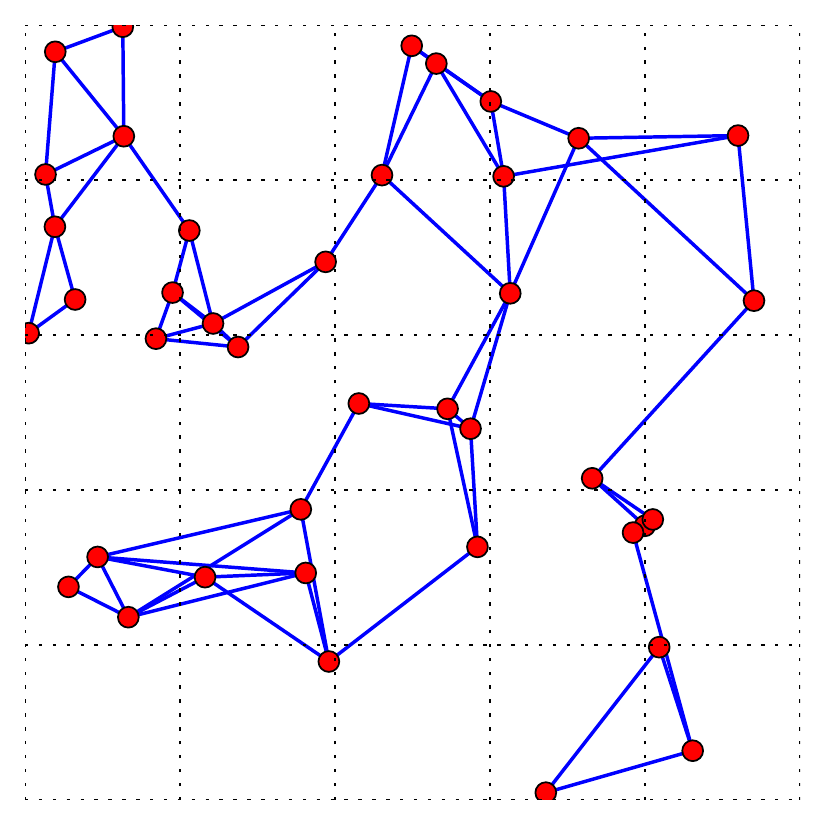}
 }
 \subfigure[40.4.A]{
   \includegraphics[width=0.27\linewidth]{data/data_40_4_1.pdf}
 }
 \subfigure[40.5]{
   \includegraphics[width=0.27\linewidth]{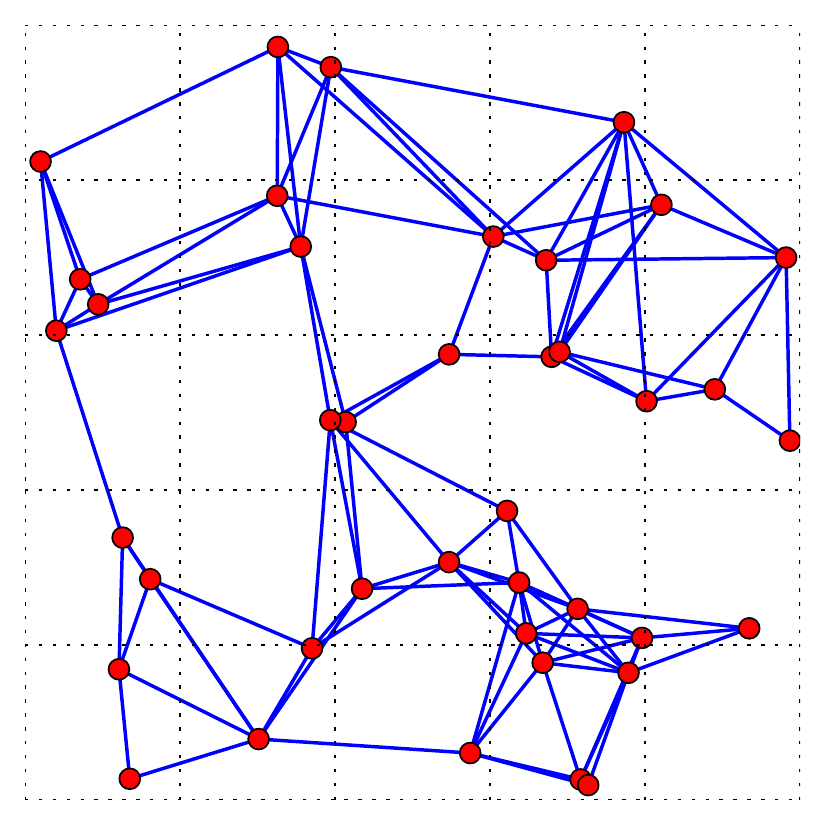}
 }
 
  \subfigure[40.6]{
   \includegraphics[width=0.27\linewidth]{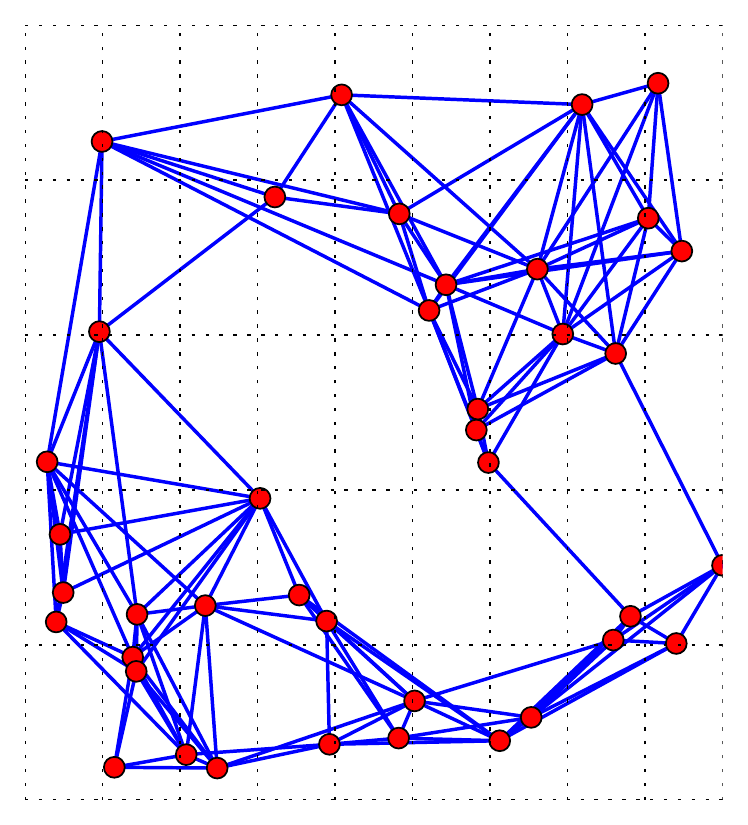}
 }
 \subfigure[40.7]{
   \includegraphics[width=0.27\linewidth]{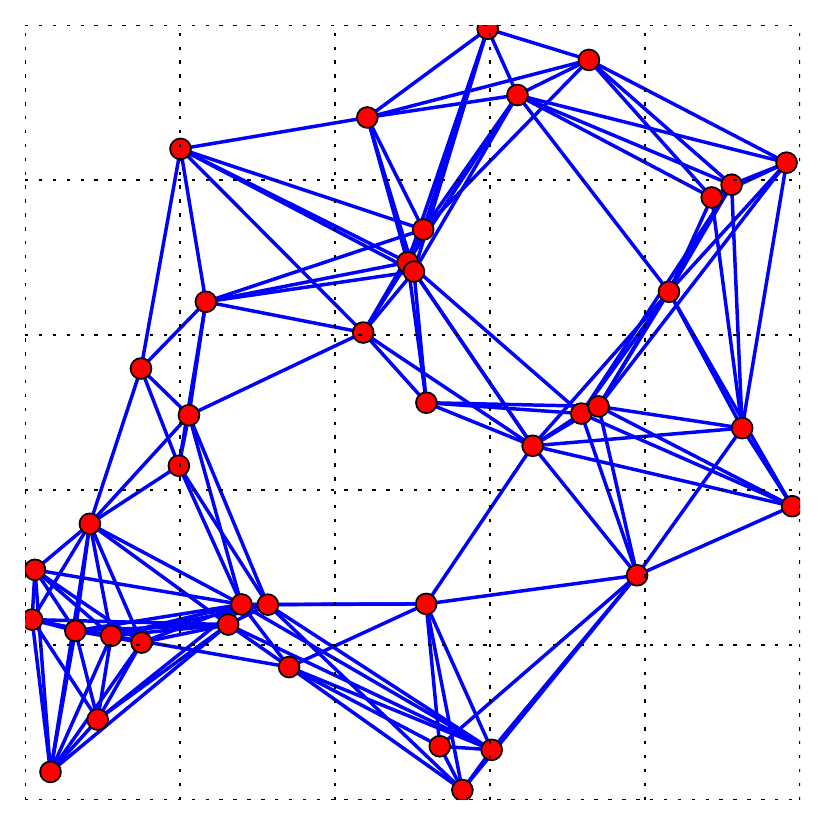}
 }
 \subfigure[40.8]{
   \includegraphics[width=0.27\linewidth]{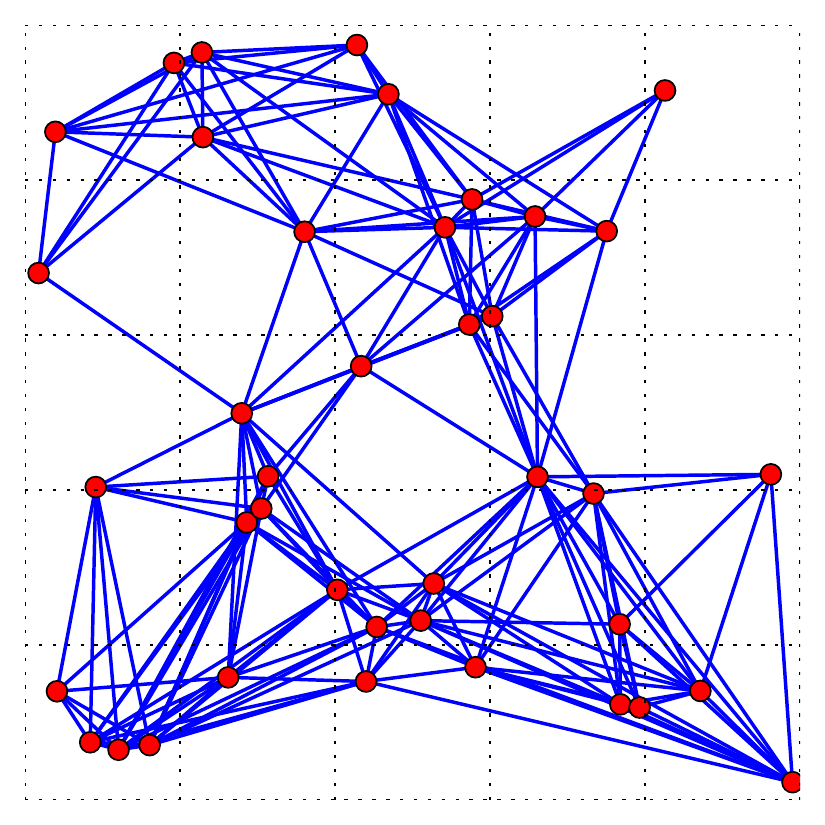}
 }
  \caption{Plot of Randomly Generated Topology Instances with Different $\lambda$}
  \label{fig:lambda}
\end{figure}

Besides the comparison of SSA for BSSP and SWES$_{(0,0)}$, we also analyze the impact of $\lambda$ on the performance of saving energy. In order to have a complete analysis on this issue, we generated 5 topologies with $\lambda\in[3,5,6,7,8]$, $|B|=40$ and conduct simulations on these instances as well as the ``40.4.A'' instance illustrated in Fig. \ref{fig:instance}. The topology instances to be considered are presented in Fig. \ref{fig:lambda}.

\begin{figure}
  \center
  \includegraphics[width=0.95\linewidth]{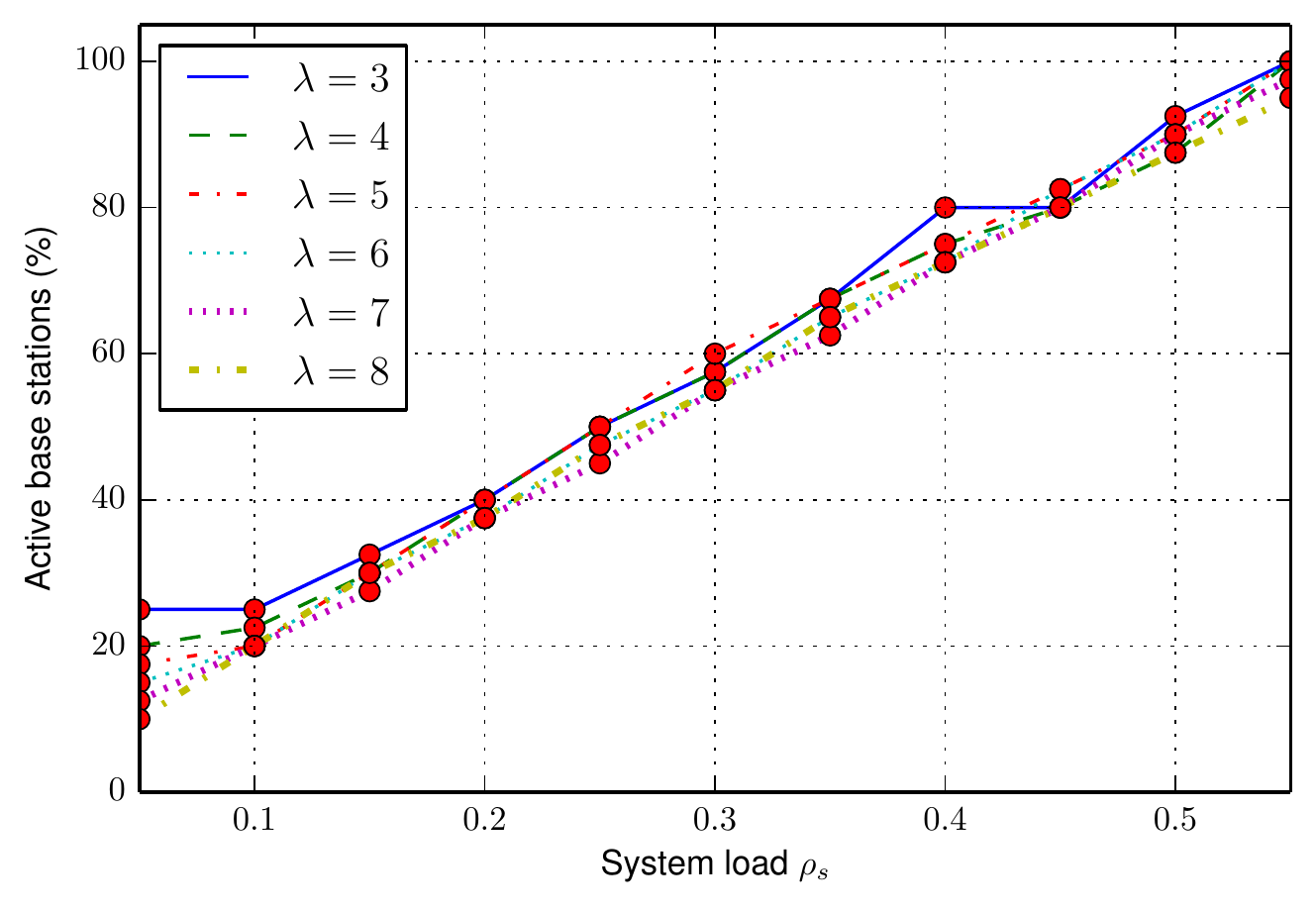}
  \caption{Simulation Results of Instances with Different $\lambda$}
  \label{fig:lambdaresult}
\end{figure}

The simulation results on these new instances with SSA for BSSP is shown in Fig. \ref{fig:lambdaresult}. From the results we can observe that although $\lambda=3$ performs slightly worse than others when the system load is high and low, the overall performance of these instances are barely distinguishable. This phenomenon suggests that increasing the neighboring links among the base stations would not significantly improve the overall energy saving performance when the system already has sufficient links. It can also be observed that the performance curves are very close to the ideal best performing curve $\rho_s/\overline{\rho_b}$. This again shows the outstanding performance our proposed algorithm.

\section{Conclusion and Future Work}\label{sec:conclusion}

In this paper we consider the possibility of saving energy in wireless cellular network by switching off the under-utilized base stations during non-peak hours. We transform the original optimization problem with constraints into an unconstrained problem with a penalty function to alleviate the design difficulty considering the constraints. Besides the new formulation, we also propose a new binary SSA to solve this optimization problem. SSA is a newly proposed general-purpose metaheuristic. It was originally devised to solve continuous optimization problem. We make several modifications to the canonical SSA to adapt it to solve BSSP. We devise two optimization schemes in SSA and make modifications on the control flow of the algorithm. We employ a series of randomly generated network topologies to test the performance of our proposed SSA for BSSP. The simulation results indicate that SSA can almost always outperform the compared algorithm. We also analyze the impact of the network density on the energy saving performance. The results imply that the performance improvement caused by the increasing network density is limited as the performance is already very close to the ideal best case scenario.

There are some potential future work. In this paper, we consider the base stations consume the same amount of energy despite different system loads and communication ranges. In reality, both factors will influence the energy consumed. This can be taken into consideration but will significantly increase the complexity of the optimization problem. Another extension is to consider the diversity of the base stations. In this work we only consider homogeneous base stations. It will be interesting to study the impact of adopting different types of base stations on the energy-saving performance.

\section*{Acknowledgement}
This research is supported in part by the University of Hong Kong Strategic Research Theme on Computation and Information.

\bibliographystyle{IEEEtran}
\bibliography{IEEEabrv,../../../bib/publications}

\begin{thebibliography}{10}
\providecommand{\url}[1]{#1}
\csname url@samestyle\endcsname
\providecommand{\newblock}{\relax}
\providecommand{\bibinfo}[2]{#2}
\providecommand{\BIBentrySTDinterwordspacing}{\spaceskip=0pt\relax}
\providecommand{\BIBentryALTinterwordstretchfactor}{4}
\providecommand{\BIBentryALTinterwordspacing}{\spaceskip=\fontdimen2\font plus
\BIBentryALTinterwordstretchfactor\fontdimen3\font minus
  \fontdimen4\font\relax}
\providecommand{\BIBforeignlanguage}[2]{{%
\expandafter\ifx\csname l@#1\endcsname\relax
\typeout{** WARNING: IEEEtran.bst: No hyphenation pattern has been}%
\typeout{** loaded for the language `#1'. Using the pattern for}%
\typeout{** the default language instead.}%
\else
\language=\csname l@#1\endcsname
\fi
#2}}
\providecommand{\BIBdecl}{\relax}
\BIBdecl

\bibitem{2013CEETAnnualReport}
{Centre for energy-efficient telecommunications}, ``{CEET} annual report
  2013,'' Bell Labs and University of Melbourne, Tech. Rep., 2013.

\bibitem{TaoAnsari2013OptimizingGreenEnergy}
H.~Tao and N.~Ansari, ``On optimizing green energy utilization for cellular
  networks with hybrid energy supplies,'' \emph{{IEEE} Trans. Wireless
  Commun.}, vol.~12, no.~8, pp. 3872--3882, 2013.

\bibitem{OhSonKrishnamachari2013DynamicBaseStation}
E.~Oh, K.~Son, and B.~Krishnamachari, ``Dynamic base station switching-on/off
  strategies for green cellular networks,'' \emph{{IEEE} Trans. Wireless
  Commun.}, vol.~12, no.~5, pp. 1536--1276, 2013.

\bibitem{CorreiaZellerBlumeFerlingJadingGodorAugerPerre2010Challengesandenabling}
L.~Correia, D.~Zeller, O.~Blume, D.~Ferling, Y.~Jading, I.~Godor, G.~Auger, and
  L.~V.~D. Perre, ``Challenges and enabling technologies for energy aware
  mobile radio networks,'' \emph{{IEEE} Commun. Mag.}, vol.~48, no.~11, pp.
  66--72, 2010.

\bibitem{MarsanChiaraviglioCiulloMeo2009OptimalEnergySavings}
M.~Marsan, L.~Chiaraviglio, D.~Ciullo, and M.~Meo, ``Optimal energy savings in
  cellular access networks,'' in \emph{Proc. IEEE International Conference on
  Communications Workshops}, Dresden, Germany, Jun. 2009, pp. 1--5.

\bibitem{YuLi2013SocialSpiderAlgorithm}
J.~J.~Q. Yu and V.~O.~K. Li, ``A social spider algorithm for global
  optimization,'' Department of Electrical and Electronic Engineering, The
  University of Hong Kong, Technical Report TR-2013-004, Oct. 2013.

\bibitem{ChiaraviglioCiulloMeoMarsan2009Energyefficientmanagement}
L.~Chiaraviglio, D.~Ciullo, M.~Meo, and M.~Marsan, ``Energy-efficient
  management of {UMTS} access networks,'' in \emph{Proc. 21st International
  Teletraffic Congress}, Paris, Fance, Sep. 2009, pp. 1--8.

\bibitem{OhKrishnamachari2010EnergySavingsthrough}
E.~Oh and B.~Krishnamachari, ``Energy savings through dynamic base station
  switching in cellular wireless access networks,'' in \emph{Proc. IEEE Global
  Telecommunications Conference}, Miami, FL, U.S., Dec. 2010, pp. 1--5.

\bibitem{FehskeRichterFettweis2009EnergyEfficiencyImprovements}
A.~Fehske, F.~Richter, and G.~Fettweis, ``Energy efficiency improvements
  through micro sites in cellular mobile radio networks,'' in \emph{Proc. IEEE
  GLOBECOM Workshops}, Honolulu, HI, U.S., Dec. 2009, pp. 1--5.

\bibitem{MarsanMeo2010Energyefficientmanagement}
M.~A. Marsan and M.~Meo, ``Energy efficient management of two cellular access
  networks,'' \emph{ACM SIGMETRICS Performance Evaluation Review}, vol.~37,
  no.~4, pp. 69--73, 2010.

\bibitem{RostFettweis2010TransmissionComputationEnergy}
P.~Rost and G.~Fettweis, ``On the transmission-computation-energy tradeoff in
  wireless and fixed networks,'' in \emph{Proc. IEEE GLOBECOM Green
  Communications Workshop}, Miami, FL, U.S., Dec. 2010, pp. 1--6.

\bibitem{KhuriBackHeitkotter1994evolutionaryapproachto}
S.~Khuri, T.~Back, and J.~Heitkotter, ``An evolutionary approach to
  combinatorial optimization problems,'' in \emph{Proc. Annual ACM Computer
  Science Conference}, Phoenix, AZ, U.S., 1994, pp. 66--73.

\bibitem{Karp1972Reducibilityamongcombinatorial}
R.~M. Karp, ``Reducibility among combinatorial problems,'' \emph{Complexity
  Computer Comput.}, vol.~40, no.~4, pp. 85--103, 1972.

\bibitem{Uetz1992Foragingstrategiesspiders}
G.~Uetz, ``Foraging strategies of spiders,'' \emph{Trends in Ecology and
  Evolution}, vol.~7, no.~5, pp. 155--159, 1992.

\end{thebibliography}

\end{document}